*Article*

# A Complex Topological Phase in C-Spin Active Matter

Alessandro Scirè[1]*

[1] University of Pavia, Department of Electrical, Computer and Biomedical Engineering
via A. Ferrata 5 – 27100 – Pavia (Italy). alessandro.scire@unipv.it

* Correspondence: alessandro.scire@unipv.it;

**Abstract**

This work explores a theoretical model of self-organization driven by the interplay of positional and orientational order in "complementary-spins" (c-spins), symbolic agents divided into two populations with contrasting positional and orientational interactions. The system, governed by a circular anisotropy parameter that makes it active, exhibits a variety of complex behaviors. For small anisotropy, uniform, stable patterns emerge. As anisotropy increases to a moderate level, the system develops robust, self-repairing topological point defects—vortex complexes characterized by orientational textures and counter-rotating c-spin loops with spin-momentum locking. These novel non-equilibrium dissipative structures are classifiable by a two-valued topological charge. Beyond a local stability threshold, active turbulence (deterministic chaos) occurs, and order is lost. A statistical analysis revealed the coexistence of a double phase transition at a critical parameter value: an "ordinary" symmetry-breaking transition and a novel topological phase transition that activates the vortex complexes. Both analytical and numerical methods were used to evaluate quantitative boundaries in the parameter space, and increasing the system size enhances the complexity of the transport loops. Due to its self-organizational properties, this model provides a new tool for understanding robustness and morphogenesis in living systems.

## 1. Introduction

Active matter [1] is a relatively new classification in soft matter physics, defined by the presence of a large number of interacting agents that consume energy to move or exert forces. This field brings together aspects from symmetry, non-equilibrium thermodynamics and topology [2]. The most extensively studied model in the field, the Vicsek model [3] published in 1995, deals with systems composed of self-propelled agents and demonstrates how local interactions can generate collective motion, displaying the emergence of complex phenomena and phase transitions. In general, the core interest in investigating organization patterns in non-equilibrium systems stems from their ability to exhibit complex processes like adaptation and self-organization, and scrutinizing these patterns provides information about how living and non-living systems achieve order, robustness, and dynamic functionality far from equilibrium. Topological states [4] are a promising tool in such sense, for being non-local and robust against disorder. Under certain conditions, active matter exhibits out-of-equilibrium structures that are both topological and dissipative, such as topological defects [5] and topologically protected edge modes (sound modes) [6]; this combination (topological and dissipative) can lead to unique and robust complex states that maintain their function despite energy loss, a phenomenon paralleled in emergent life forms [7].

Recently, topological phases in active matter have been found to influence the mechanics and organization of living tissues: an active nematic [8] state can emerge in epithelial cell layers, where cells collectively align and generate active forces from their internal cytoskeletons, and spontaneously generated topological defects move and annihilate in pairs. These defects, which emerge from the collective, self-organized movement of cells, serve as crucial organizing centers for biological processes. By modeling epithelial layers as active nematic liquid crystals, researchers have demonstrated that the stresses generated by these defects can dictate cell fate and drive large-scale tissue behaviors like extrusion, migration, and morphogenesis [9]. In the Hydra organism, the precise arrangement of topological defects is utilized to position its limbs, mouth, and foot [10].

Aside from the aspects concerning active nematics (mostly developed in the frame of the modified Navier-Stokes equation), understanding how self-propelled units coordinate their movement, orientation, or internal oscillations to produce coherent large-scale behavior also falls within the frame of collective synchronization. Arthur Winfree pioneered the use of mathematical models to describe biological clocks, non-equilibrium systems that consume energy (ATP) to maintain rhythm and synchronize with environmental cues. His Winfree model [11] provided an early, complex framework for synchronizing diverse oscillators influencing Yoshiki Kuramoto, who developed the more analytically tractable Kuramoto model [12] establishing the foundational theories for collective synchronization in biological systems. The transition to synchronization in the Kuramoto model is a classic example of spontaneous symmetry-breaking phase transition in a non-equilibrium system. Still, the Kuramoto model can be modified to exhibit topological phase transition: When it is placed on a spatial lattice (like a two-dimensional grid), the relative positioning is fixed, but the spatial pattern of the synchronized phases exhibits features of positional/topological order related to the Berezinskii-Kosterlitz-Thouless transition and defect binding [13].

However, the (highly successful) Kuramoto model – including its numerous variations [14] – describes the synchronization of phases or internal states, not the collective motility characteristic of active matter agents. The standard Kuramoto model and most of its traditional variations deal only with orientational order (phase synchronization); it focuses on the internal phases of oscillators at fixed spatial locations, without explicitly modeling their physical motion in space.

A significant recent variation – the 'swarmalator' [15] model – explicitly links the internal Kuramoto phase dynamics to the spatial dynamics of mobile agents. Such model exhibits both phase and spatial order, such as states where particles cluster in space and synchronize their phase simultaneously. In specific theoretical variations, particularly in the hydrodynamic (continuum) limit or models with non-reciprocal forces, swarmalator models exhibit states of a topological nature [16].

However, despite exciting progress in connecting collective synchronization to symmetry-breaking and topological phase transitions, significant aspects of topological protection in space-time synchronization of active agents remain largely unexplored. In the absence of a general theory, research relies on the creation of new paradigms.

This work contains a quantitative analysis of the organizational properties of a recently introduced [17][18] theoretical model for the interplay of positional and orientational (phase) order. The model consists of many spin agents able to rotate and move in 2D, and divided into two sub-populations. Those spins interact in a complementary scheme: when belonging to the same population, spins repel each other positionally and attract orientationally; when belonging to different populations they do the opposite. Hence the two spin sub-populations interact in a complementary way, so I have named those units *complementary-spins* or *c-spins* in short. The interaction decays with the distance. The system is given two parameters, the total number of elements (equally divided between the two populations) and a control parameter that splits the natural rotational frequencies of the two populations, and induce a locking-to-unlocking phase transition. Also, the control parameter makes the system active, i.e. make the units capable of orientational motion, which, due to variable coupling, translates to spatial self-propulsion.

The system displays the coexistence of a double phase transition at the same critical parameter value: an ordinary symmetry-breaking phase transition associated with collective synchronization and a novel topological phase transition that activates robust vortex complexes. These complexes exhibit properties characteristic of a topological phase, including non-locality, robustness, effective dissipationless transport, spin-momentum locking and topological invariants. I have investigated these transitions using collective parameters and derived analytical formulas (either by calculation or by numerical fitting) to generate a first portrait of the phase diagram in the parameter space.

When increasing the system size the collective organization shows a significant increment of complexity: the single vortex appearing in small systems is replaced by a flexible morphogenetic flow of complex spin-momentum locked states, connecting this work with previous findings [17, 18]. Many different scenarios unfold at that point, which will be the subject of future research.

The manuscript is organized as follows: Section 2 describes the model. Section 3 presents the main findings of this work, subdivided into four subsections: Subsection 3.1 provides a simple derivation of the threshold value of the control parameter for the order-disorder transition. Subsection 3.2 first presents examples of possible collective dynamics for a small, but statistically significant, ensemble, and then focuses on the statistical properties as the control parameter is varied from below to above the threshold. Subsection 3.3 discusses the phase diagram in the parameter plane N Vs $\Delta$. Subsection 3.4 provides illustrative examples and comments concerning the complex collective dynamics that occur in larger systems. Section 4 is dedicated to summarizing and concluding the manuscript emphasizing the reasons of interest of this work.

## 2. The Complementarity Model in the c-spin representation

The model under analysis has already been considered in preliminary works [17, 18]. It consists of many agents endowed with spatial (2D) and rotational degrees of freedom and they are divided into two sub-populations. Differently from previous publications [17, 18, 19], in this work each population is represented by a distinct color (blue and red), and the positional/rotational dynamics is represented by mobile symbolic spins. The system is thus depicted as many two-color rotating arrows moving in a plane, as sketched in Fig.1 panel (a). This representation proved more effective than the previous one for visualizing the morphology of the emerging orientational patterns (defects) and waves, as further elucidated in the manuscript.

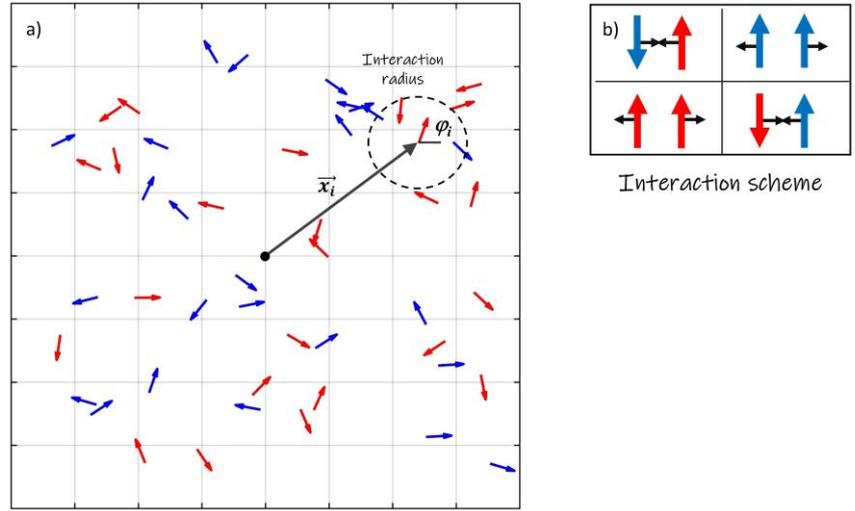

**Figure 1**. Sketch of the Complementary-Spin Model. a) The model assigns both a position and an angle in [0,2π] to each spin. b) Interaction forces are complementary based on color: same-colored spins repel positionally and attract orientationally, while different-colored spins do the opposite. All interaction strengths decrease with increasing positional distance.

Each spin is given a positional ($\vec{x}_i$) and an angular/orientational ($\varphi_i$) degree of freedom in [0, 2π], with $i = 1…N$, being $N$ the total number of spins. I assume as many blue spins as red ones. Those spins interact by a *complementary scheme* as sketched in Fig.1 panel (b), i.e. spins of the same color repel positionally and attract orientationally, whereas spins of different color do the opposite. The interaction strength decays with the positional distance hence it is a short-range interaction model.

The equations of motion for each *i-th* c-spin are

$$\begin{cases} \dot{\vec{x}}_i = \sum_{j=1}^{N} \nabla_i \, W(|\vec{x}_i - \vec{x}_j|) \cos(\varphi_i - \varphi_j), & (1) \\ \dot{\varphi}_i = \gamma_i \Delta + \sum_{j=1}^{N} \gamma_i \gamma_j \, W(|\vec{x}_i - \vec{x}_j|) \sin(\varphi_i - \varphi_j), & (2) \end{cases}$$

where $\nabla_i$ means differentiation respect to the *i-th* direction, $\vec{x}_i$ and $\varphi_i$ are respectively the spatial (in $\mathbb{R}^2$) and angular (in $S^1$) coordinates of the *i-th* unit, *i = 1,…N*, and $|\vec{x}_i - \vec{x}_j|$ is the Euclidean distance in $\mathbb{R}^2$. Since the system (1)-(2) is based on a complementary type of interaction, I have named *c-spins* (meaning complementary-spins) the two-type units of the systems. The functional wells (that make the interactions *local* with characteristic length L = 1) are chosen as exponential:

$$W(|\vec{x}_i - \vec{x}_j|) = -e^{-|\vec{x}_i - \vec{x}_j|^2}. \qquad (3)$$

The specific choice of an exponential decay does not affect qualitatively the results. The two-valued coefficients $\gamma_i$ define the *i-th* oscillator color, i.e. $\gamma_i$ = +1 if the oscillator is a red or $\gamma_i$ = −1 if is blue. Mathematically, the model (1)-(2) is a many-body dissipative and non-linear dynamical system, and $\Delta$ is a control parameter. The effect of $\Delta$ in Eq. (2) is to split the natural frequencies of red and blue spins, providing a rotational speed term that is clockwise for red spins and counterclockwise for blue spins; in a physical system $\Delta$ would be called a *circular anisotropy*, hence I will often refer to $\Delta$ as 'anisotropy' in the rest of the manuscript. Collectively, the parameter $\Delta$ triggers an order-to-disorder phase transition in the system upon overcoming a specific locking/unlocking threshold, driving the system to deterministic chaos. The parameter $\Delta$ makes in fact the system *active*, because it provides the spins a phase drift which, due to variables coupling, can indirectly activate positional dynamics. The term $\cos(\phi_i - \phi_j)$ acts as "spin-orbit" coupling—it links the spatial motion to the internal orientation state; this is crucial for topological structure formation as elucidated further in the manuscript. The $\sin(\phi_i - \phi_j)$ term acts as the synchronization torque, a hallmark of the Kuramoto model.

### 3. Results

#### 3.1. Dipolar solutions and local instability threshold

According to the interaction scheme reported in panel (b) of Fig.1 the c-spins tend to organize in antiparallel pairs of red-blue spins (I have named those pairs *c-dipoles*) i.e. forming an angle $\varphi = \pi$, locked in relative orientation and relative position. This is strictly true only when $\Delta = 0$, in fact, from (1)-(2) it can be shown that the relative angle $\varphi$ between two c-spins in a c-dipole (far from other interactions) satisfies the following equation:

$$\dot{\varphi} = 2\Delta + sin\,\varphi, \qquad (4)$$

i.e. the well-known Adler equation [20], which is closely related to the individual oscillator dynamics within the broader Kuramoto model framework discussed earlier. The locking threshold is $\Delta_{th}$ = ½, below this threshold – which determines a local transition from stable to unstable dynamics as pictured in Fig. 2 – the two c-spins are locked to the same spatial position, and to a relative angle equal to

$$\varphi = \pi - arcsin(2\Delta), \qquad (5)$$

that reverts to φ = π when Δ = 0. If |Δ| > ½ the two c-spins unlock via saddle-node bifurcation and display a dynamic state in both position and orientation. Without loss of generality, in the following I will consider positive values for Δ.

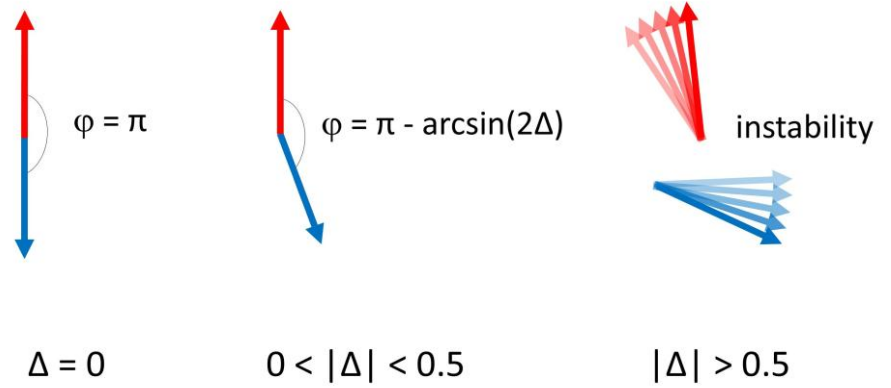

**Figure 2.** Illustrates the phase dynamics and locking behavior of a c-dipole. C-spins typically form antiparallel pairs called c-dipoles, which align at an angle of π when the detuning parameter Δ=0. The Adler equation, which governs oscillator locking, describes the orientation and position of these pairs and predicts a locking threshold at $\Delta_{th}$ = ½. Below threshold the c-spins are phase-locked at a specific relative angle and to the same spatial position; above threshold they unlock through a saddle-node bifurcation and become dynamic.

*3.2. Collective behavior of a small ensemble*

In this subsection I report on numerical simulations concerning an ensemble of N = 80 c-spins for increasing values of the control parameter Δ.

*Weak anisotropy*

I have numerically integrated, by means of a standard method, Eqs. (1)-(2) for N = 80 c-spins with Δ = 0.1, i.e. well inside the stable locking range Δ < ½. The initial conditions are random spin angles with uniform distribution between 0 and 2π, and random spin positions uniformly distributed in a square box; the size of the box is chosen equal to $\sqrt{N}$ so that the c-spin density inside the box is equal to 1. As discussed in previous works [17, 18, 19] this density choice is not critical, but if the units are too sparse, they tend to separate in non-interacting clusters, whereas if too compressed, the collective flow discussed later are hindered and struggle to emerge.

The resulting spatiotemporal dynamics is shown in Movie S1 and the regime configuration is reported in Fig. 3.

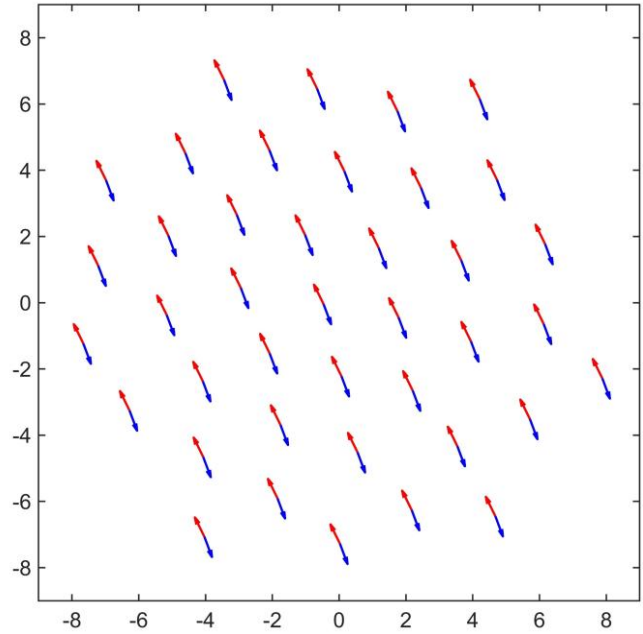

**Figure 3.** A type ① solution: a pattern of c-dipoles possessing both positional and orientational order. This pattern represents an equilibrium state resulting from the spontaneous breaking of continuous circular symmetry, where c-dipoles are aligned at a specific angle, randomly determined by initial conditions.

The pattern reported in Fig. 3 is made of (quasi) regularly spaced c-dipoles, each of which fulfils the local stability properties of the single c-dipole (see Fig. 2b). The pattern of Fig. 3 represents an equilibrium state associated to the spontaneous breaking of a continuous circular symmetry, like in ferromagnetic systems or in collective synchronization, the c-dipoles are forced to align at a specific angle, randomly determined by the initial conditions. This kind of pattern is referred to hereafter as a Type ① solution. The spatial regularity is due to a weak repulsive dipole-dipole force, which is proportional to $\Delta^2$ and pushes the pattern to expand. The expansion rate is logarithmic; consequently, for practical purposes, it vanishes when it attains approximately three times the interaction length.

*Moderate anisotropy*

Increasing $\Delta$ without crossing the local stability threshold, a different collective behavior shows up, bistable with type ①. The Movie S2 shows the spatiotemporal evolution of $N = 80$ c-spins with $\Delta = 0.2$. After the transient has expired, the movie displays the appearance of a topological vortex complex exhib-iting a peculiar two-fold structure: a static texture, where the c-dipoles curl around a central core, surrounding a dynamic state made by two trains of equidistant counterpropagating c-spins that insist on the same loop trajectory. The distance between adjacent c-spins of the same color in the loop is equal to 1 in average, with very small deviation; the orientational and the positional motions are synchronized in a Spin-Momentum locked state, i.e. one full rotation in spin correspond to one full rotation in the space loop. Surprisingly, the red and blue c-spin trains slide over each other "frictionless", despite the value of Δ (smaller than the locking threshold) would require that c-spins of different colors attract and lock in the same position. I have called that vortex complex (static texture + dynamic loop) a type ② solution.

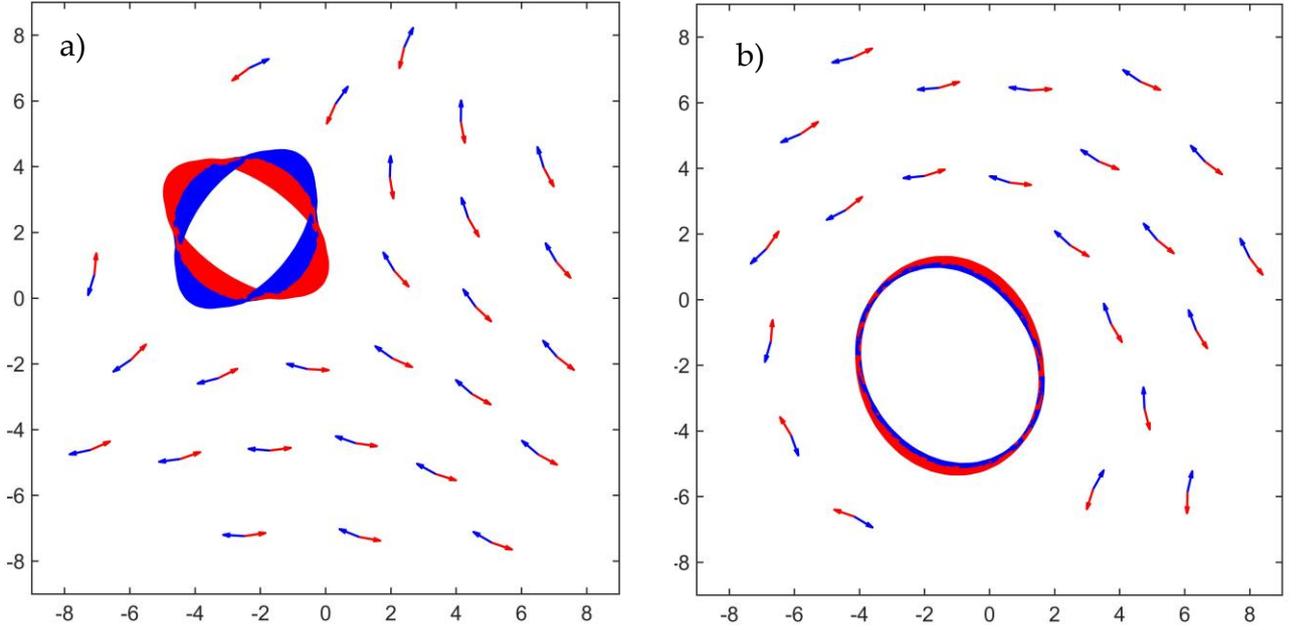

**Figure 4**. Illustrates persistent features by superimposing video frames over time for N = 80 and Δ = 0.2. (a) A stable antivortex complex with a topological charge Q = - 1; (b) A stable vortex complex with topological charge Q = 1, emerging for the same parameter values.

Emergent Type ② solutions are topological point defects driven by phase singularity, and represent a dynamic balance between alignment interactions and "spin-orbit" coupling. They appear in two forms, for the same parameter values, as illustrated in a superimposition of video frames (akin to a long-exposure photograph, shown in Figure 4). These forms are classified (see Table 1) based on their winding number (or topological charge, Q), which measures the net rotation of the c-dipoles orientation along a counterclockwise (CCW) closed path around the defect core. These complexes constitute a whole new class of non-equilibrium dissipative states.

| Table 1. Vortex vs. Antivortex Classification ||||
|---|---|---|---|
| Complex Type | Description | Winding Number Q | Figure |
| Antivortex | Units rotate once CCW around the center | -1 | 4a |
| Vortex | Units rotate once CW around the center | +1 | 4b |

*Diversity of attractors and topological invariants*

The attractors emerging from repeated simulations of Eqs. (1)-(2), within the locking region and for a given system size (N = 80), show a wide range of morphological diversity; still, all them can be ascribed to one out of three distinct types based on their topological properties: type ① solutions (equilibrium states with Q = 0), type ② vortex (Fig. 5 a, d and e) with Q = 1, or type ② antivortex (Fig. 5 b and f) with Q = -1. The emerging type ② solutions display a high degree of variability in terms of: the number of elements involved in the internal loop; the phase singularity position in the plane; the vorticity and the velocity of the transport motion (momentum); and the spin rotation velocity in the loops. Rarely, a double loop with a daughter loop branching off from the main formation (see Fig.5d) appear, providing evidence of the system's capacity for morphological complexity, as discussed in a later section; however, even that "strange" configuration belongs to the Q = +1 topological class.

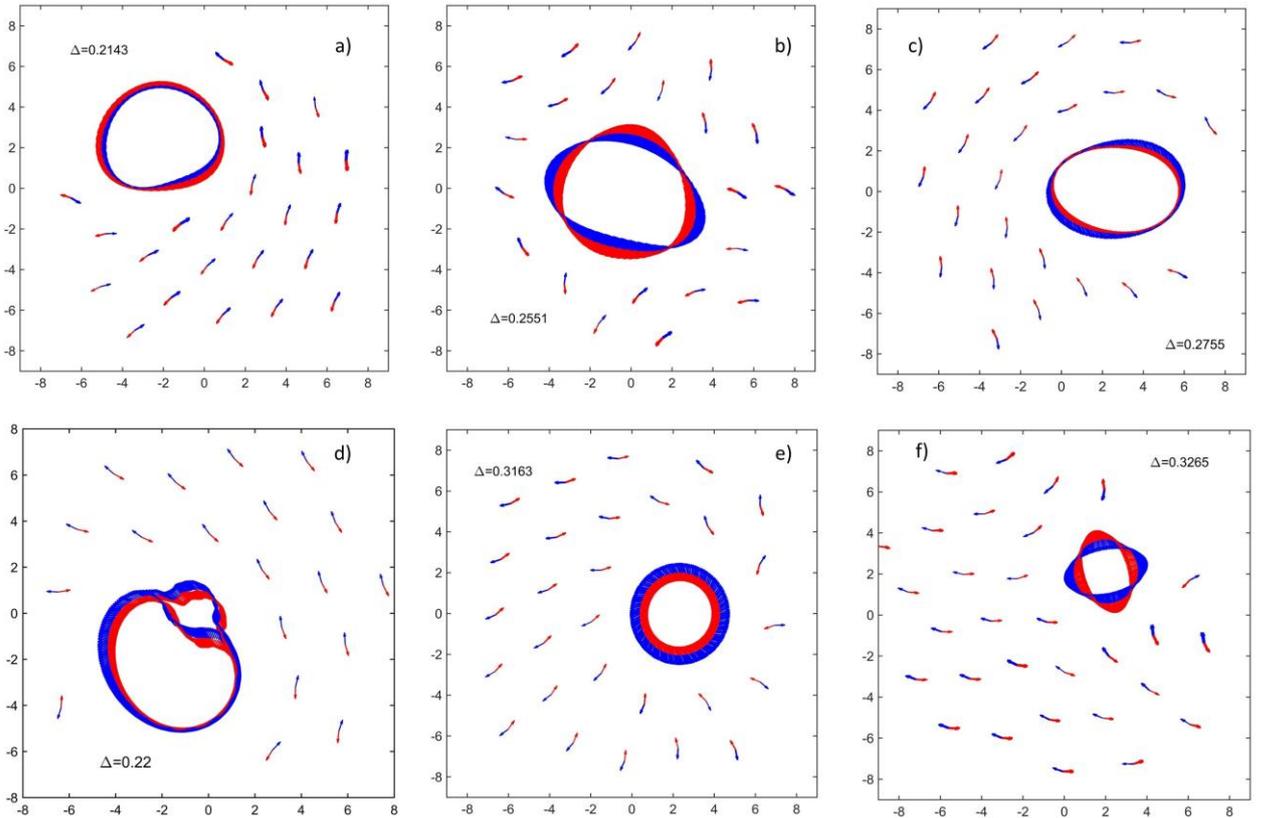

**Figure 5**. Illustrates the diversity of emerging attractors obtained from repeated numerical simulations using different values of Δ < Δth, N = 80. The system exhibits multistability, settling into either equilibrium type ① solutions or type ② solutions, which include both vortices (panels a, d, e) and antivortices (panels b, f) complexes. Type ② solutions display high variability in the number of elements, position, vorticity, momentum and spin velocity.

Above the locking threshold, numerical simulations for Δ > ½ show spatiotemporal chaotic dynamics in both position and orientation, a form of active turbulence, a global consequence of the local unlocking transition. I call type ③ this kind of supercritical patterns. Increasing further Δ the behavior becomes more erratic and finally no trace of order is left.

*Robustness against perturbations*

The primary implication of a topological invariant is robustness. In a topological state, the value of the invariant remains unchanged under continuous deformations, minor perturbations, or the high degree of diversity in spatial arrangements. In fact, the complementary regime structures obtained in the previous subsection, made of vortex and antivortex complexes, proved to be highly resilient with respect to different kinds of perturbations. First, the structures persist stable under the addition of Langevin white noise terms to Eqs. (1)-(2) even at high levels of noise. Second, I perturbed the structure emerged in Movie S2 in the following way: after the structure is formed, I removed one red spin from a c-dipole forming the external texture and place into the vortex, close to the center, the result in reported in [Movie S3](Movie S3). The displaced red element is quickly absorbed in the vortex and, after a short time, a red spin is expelled from the internal loop so that it can couple back with the blue spin left uncoupled and reform a c-dipole to restore the original topology. This observed self-repairing capability is a direct consequence of the non-local topological protection inherent to these complexes.

The reported results suggest that a relatively small ensemble of N = 80 c-spin shows a rich multi-stable scenario that requires a systematic analysis. Moreover, the type ② patterns are non-local solutions that cannot be understood in terms of a local stability analysis, they are therefore hard to tackle analytically moving from Eqs. (1)-(2), so I have relied on a statistic analysis based on extensive numerical simulations.

*3.3. Statistical analysis*

Considering the system (1)-(2), I have performed reiterated numerical simulations and Δ spanning from 0 to 1, with different realizations for each value of Δ, starting from random initial conditions with the same statistical properties: random positions in a box with spin density equal to 1 and uniform random orientations in [0, 2π]. I have taken as a reference a collective parameter: the total time averaged kinetic energy $T_{av}$

$$T_{av} = \langle \frac{1}{N} \sum_{i=1}^{N} \dot{x}_i^2 \rangle \ , \qquad (7)$$

that account for spatial fluctuations; $\langle \cdot \rangle$ stands for temporal average. The results for $T_{av}$, versus $\Delta$ are shown in Fig. 6a.

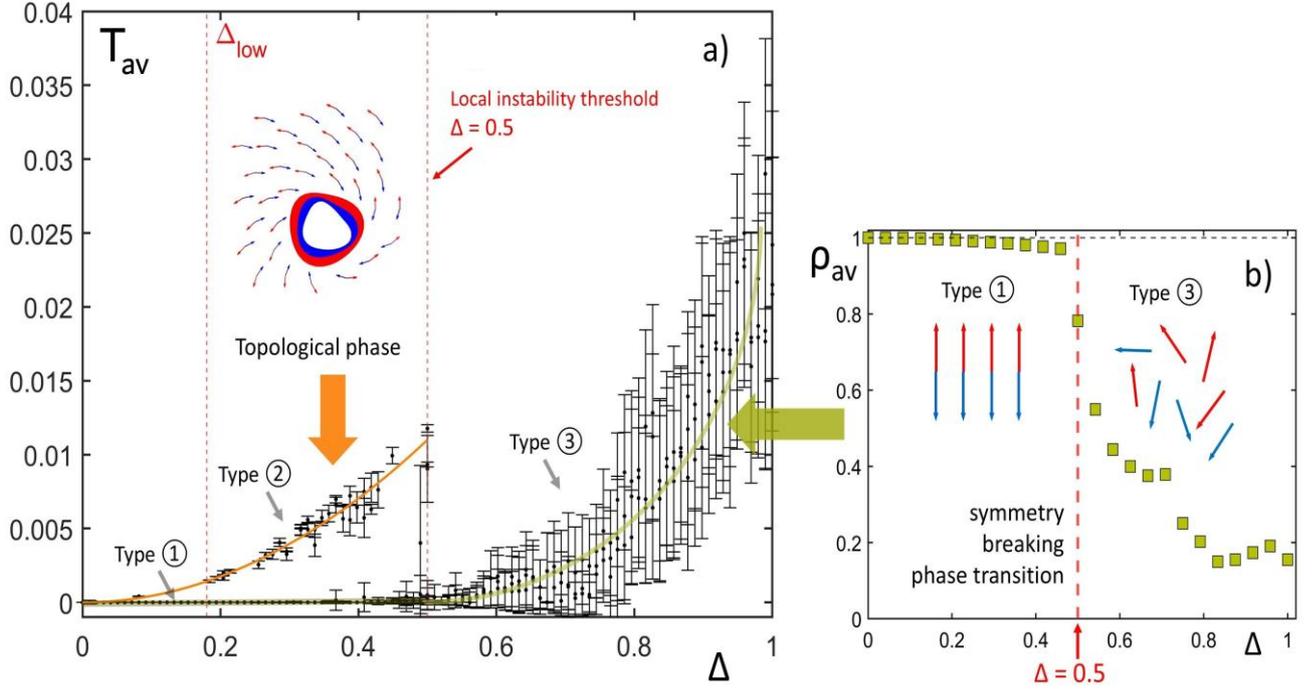

**Figure 6**. Panel (a) shows the time-averaged kinetic energy $T_{av}$ as a function of the control parameter $\Delta$. The plot reveals distinct behavioral regimes: a small $\Delta$ regime (type ① patterns) with vanishing $T_{av}$; a bistable intermediate regime $\Delta_{low} < \Delta < \Delta_{th}$ ($\Delta_{th} = 0.5$ is the locking instability threshold) featuring a lower branch of type ① patterns and an upper branch of type ② patterns (example in the inset); and a supercritical turbulent, chaotic regime $\Delta > \Delta_{th}$ where all order is lost. Panel (b) displays the time-averaged absolute value of the Kuramoto order parameter ($\rho_{av}$) as a function of $\Delta$. The panel illustrates a Kuramoto-like phase transition behavior, specifically the transition from type ① (synchronization, $\rho_{av} = 1$) to type ③ (desynchronization/turbulence, $\rho_{av} \to 0$), obtained from initial conditions near the type ① basin of attraction.

Panel (a) of Fig. 6, shows $T_{av} \approx 0$ for low values of $\Delta$, approximately $0 < \Delta < \Delta_{low}$ with $\Delta_{low} \approx 0.18$. In this region the equilibrium type ① patterns appear, stably. For $\Delta_{low} < \Delta < \Delta_{th}$ the values of $T_{av}$ separates in two branches. The lower branch is a continuation of the type ① pattern branch from the previous interval, and is bistable with a second branch containing the twofold type ② patterns discussed in the previous subsection. The upper branch follows approximately a $\Delta^2$ dependence (orange fitting line in Fig.6a) and contains an infinite number of different attractors, separated in two families classified via Q). For $\Delta > \Delta_{th}$, local instabilities take places as predicted by Eq. (4), all the c-dipoles unlock destroying both type ① and ② attractors, and chaotic dynamics (type ③ attractors) appear, increasingly turbulent as $\Delta$ increases further. Close after $\Delta_{th}$ small vortices may still be present, together with other dynamic patterns and floating spins, whereas for high values of $\Delta$ any trace of positional and orientational order is finally lost.

The transition from type ① to type ③ patterns is understood in terms of the breaking of a local circular symmetry, parallel to the transition to a ferromagnetic state, or to the Kuramoto transition to collective synchronization. It is a "classical" spontaneous symmetry breaking phase transition of the Landau type. Indeed, it can be described by means of the Kuramoto order parameter, modified as in previous works [17, 18, 19]

$$\rho e^{i\theta} = \frac{1}{N}\sum_{k=1}^{N} \gamma_k e^{i\varphi_k}. \qquad (8)$$

The time averaged absolute value $\rho_{av} = \langle \rho \rangle$ is bounded between 0 and 1, where 1 means total phase synchronization and 0 total phase desynchronization. Panel (b) of Fig. 6 shows the typical Kuramoto-like phase transition behavior when $\rho_{av}$ is computed versus $\Delta$. Fig. 6b concerns the transition type ① to type ③, obtained by making the system start from initial conditions close to the basin of attraction of type ① solution.

It is worth noticing that the point $\Delta = \Delta_{th}$ organizes a double phase transition, a Kuramoto-like and a topological one. As already mentioned, the emergence of type ② attractors is not related to a spontaneous symmetry breaking of a local symmetry, it displays the characteristic features of a topological phase. Such features are: the robustness and the non-local character, the topological invariants, the dissipationless transport and the spin-momentum locking. Those type ② patterns contain features surprisingly parallel to those of quantum states of matter. Specifically, the patterns exhibit features that parallel the spin helical transport regime found in topological insulators [21], showcasing robust, counterpropagating dissipationless flow, also found here, within a dynamical non-equilibrium system, for the first time to the best of my knowledge. However, together with similarities with other systems, the c-spin model shows consistent peculiarities. In particular, the scenario grows increasingly complex and peculiar when more c-spins are considered.

### 3.4. *Larger systems*

In this subsection, the evolution of type ② solutions is examined as the system size N increases by means of a few examples. As a general remark, the single loop patterns observed in smaller systems are replaced by a flexible, morphogenetic flow that generates dynamic compartments and a progressively more intricate spin-momentum locked flows as N increases, displaying both long-range order and topological protection. Figure 7 illustrates the progression from: panel (a) for N=200 (simple figure-eight shape, see [Movie S4](Movie S4) for the regime spatiotemporal dynamics) to panel (b) for N=500 (more convoluted boundary loops, see [Movie S5](Movie S5)) to panel (c) for N=1000 (highly intricate, compartmentalized network, see [Movie S6](Movie S6)). A detailed analysis of the N = 1000, $\Delta = 0.01$ case is provided in reference [18]. The spin-momentum relationship is still a kind of synchronization, but more intricate respect to the simple 1:1 locking displayed by smaller systems, and it percolates the whole system size.

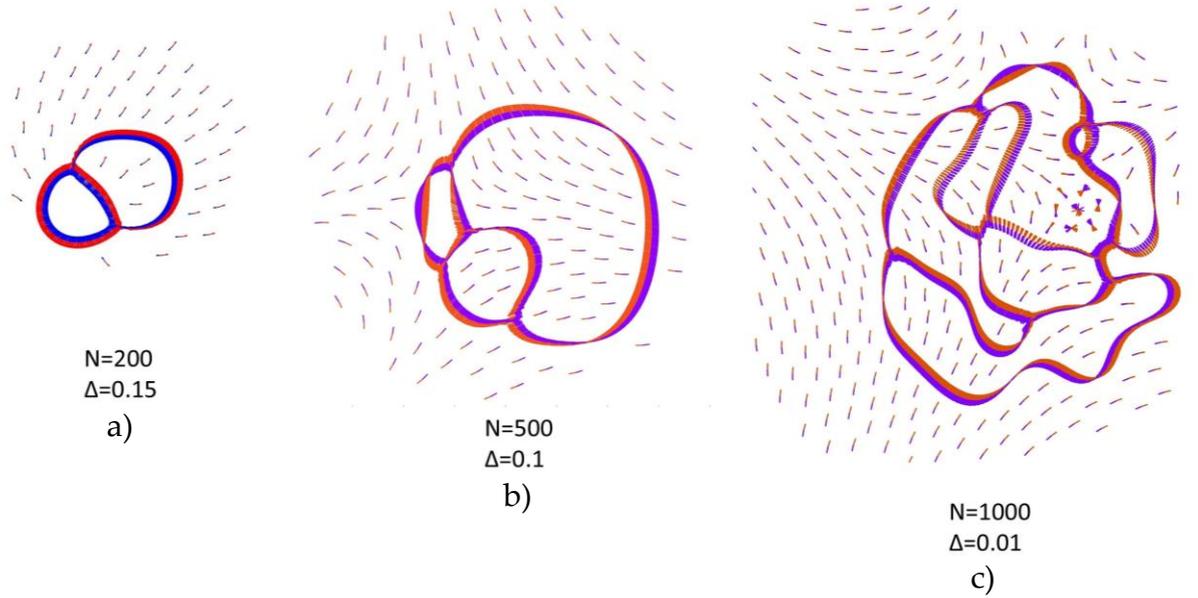

**Figure 7**: Sketch illustrating the increase in complexity of type ② solutions as system size N increases. The single vortex observed in small systems is replaced by a progressively complex, morphogenetic flow that forms increasingly intricate spin-momentum locked networks.

Fig. 7c also shows instabilities in the c-dipole textures, witnessing a weakening of the topological protection. Indeed, in [18] it is reported that such network is only marginally stable and has a finite lifetime; a possible reason for it is unveiled in the next section.

The big sizes are numerically hard to tackle, the flow stabilizes after a long, non-exponential "glassy" transient with power-low frequency spectrum [18]. Many scenarios open, that will be the object of future research.

*3.3 Phase diagram*

By exploring different system sizes, ranging from N = 80 to N = 1000, I have observed that the values of $\Delta$ required to obtain type ② solutions decrease as the system size N increases. To highlight the transition between the stability region of type ① and the bistable region containing both type ① and ②, I performed extensive numerical simulations across the (N, $\Delta$) parameter plane. The result is reported in Fig. 8, where white dots indicate the presence of a type ① uniform solution and black dots denote the emergence of type ② vortex complexes. The analysis reveals that as N increases, the transition value of $\Delta$ (previously termed $\Delta_{low}$) decreases following the empirical scaling relationship

$$\Delta_{low}(N) = \frac{a}{N}, \qquad (9)$$

with $a$ is a fitting parameter. Repeated simulations consistently yielded similar results, and data from single runs at higher values of remained consistent with the scaling behavior described by Eq. (9). As an addendum to previous results, what was called "network death" in [18] can understood in the light of Eq. (9). In [18], the emergence of a complex spin-momentum locked network with N = 1000 and $\Delta$ = 0.01 (sketched Fig.7c) was reported. The network revealed not to be stable in the very long run. In fact, Eq. (9) gives $\Delta_{low}$(N=1000) = 0.012 as the lower stability threshold for those kind of solutions; so, being close to threshold but outside the stability region, the network was marginally stable and did not endure.

Since the density is fixed, N (the number of c-spins) also represents the area of the squared box where the dynamics initiate (see inset in Fig. 8). Consequently, the significant value is the product of the anisotropy $\Delta$ and the system size N. This product N$\Delta_{low}$ effectively acts as a minimum 'quantum' necessary to assist the formation of the topological phase, and, since $\Delta$ cannot exceed $\Delta_{th}$= 0.5 which would activate chaotic behavior, the minimum size for having vortex/antivortex complexes is N = 24.

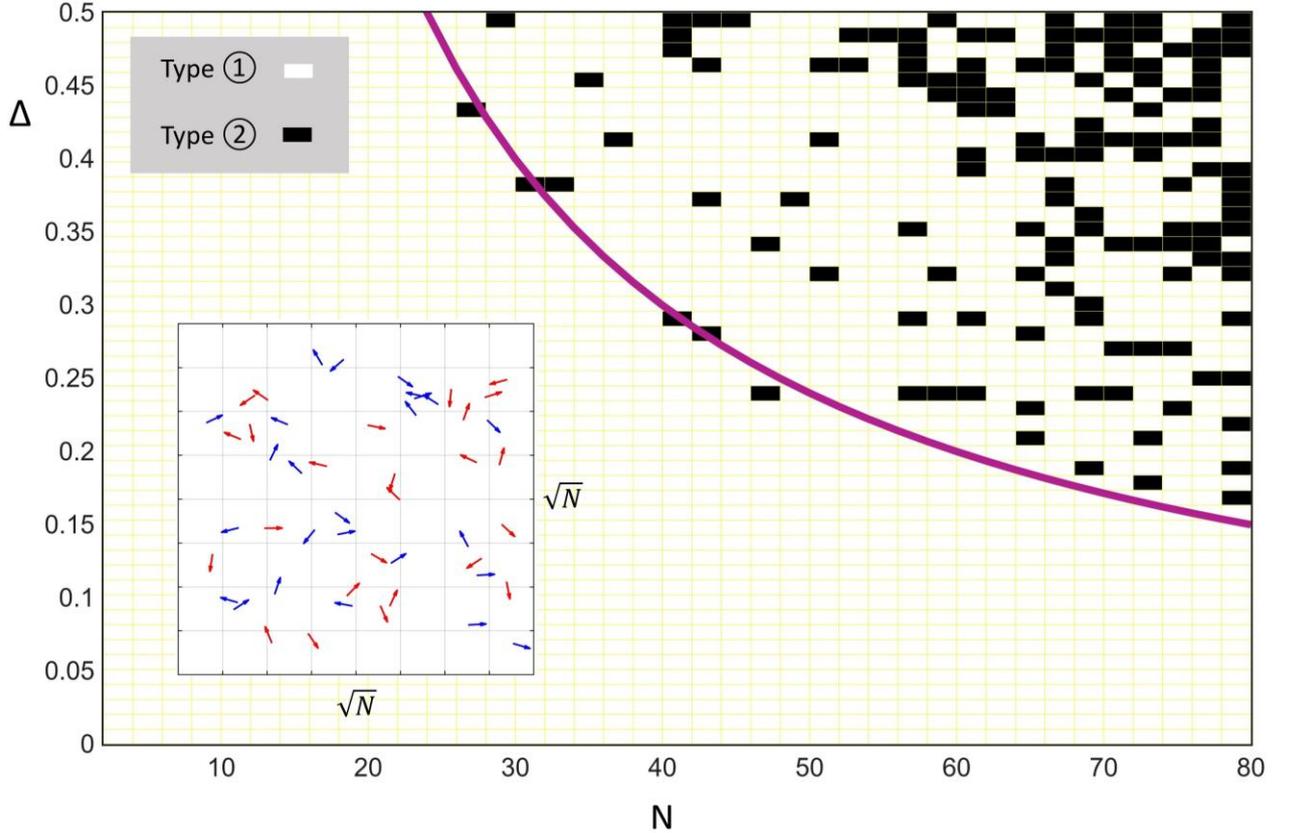

**Figure 8**: Phase diagram in the (N, $\Delta$) parameter space, illustrating the transition between different solution types. White dots indicate regions where stable type ① uniform solutions appeared, while black dots denote the emergence of type ② complexes. The boundary value $\Delta_{low}$ is fitted with $\Delta_{low}(N) = \frac{a}{N}$ (purple curve); $a$ = 12.

The number N = 24 appears to be the smallest number of c-spins that can structurally support the complex arrangement required for a stable, self-organized vortex or antivortex composite, it represents the point where the behavior transitions from single-unit interactions to collective organization: the onset of collective behavior: Below N = 24, the local interactions dominate, and complex collective structures cannot "nucleate" effectively.

## 4. Discussion

The presented model offers a novel framework for investigating the reciprocal effects of synchronization dynamics and physical space organization. This model features two populations of symbolic agents, termed complementary-spins (c-spins), which exhibit antagonistic positional and orientational interactions, introducing crucial frustration into the system. The red and blue spins prefer to rotate in opposite directions due to circular anisotropy, but the coupling term tries to synchronize them.

The system shows topological characteristics. The coupling of the orientational and positional degrees of freedom facilitates the existence of point defects characterized by integer winding numbers, associated to phase singularities. Indeed, composite patterns of topological orientational textures and spin-momentum locked loops emerge. The textures contain topological charge and the loops exhibit ballistic flow when local properties would predict high dissipation. The loops and the textures are complementary parts of the same phenomenon, a composite topological defect. The compound (texture + loop) patterns are non-local, self-organized vortex complexes, extremely robust to noise or perturbations. These features are distinctive hallmarks of a topological phase, and these patterns can be classified as topological dissipative structures. The spin-momentum locking displayed by the loops is due to the dynamic coupling between orientational and positional degrees of freedom present in the model. The ordered loops behave ballistically because the system has found a stable, low-energy configuration that perfectly satisfies the "spin-orbit" coupling constraints, a state that is topologically protected—it cannot simply decay via friction or dissipation into a uniform state without a large energy input or annihilation with another defect of opposite charge. This stability makes the flow behave like a coherent, "dissipationless" collective excitation or soliton.

I have investigated the vortex complexes by means of illustrative movies and by numerically evaluating two collective parameters concerning spatial activity (the average kinetic energy) and phase synchronization (a modified Kuramoto order parameter). The system shows the coexistence of a twofold phase transition taking place at a critical value of the control parameter: an "ordinary" symmetry-breaking phase transition associated with collective synchronization (as in the Kuramoto model) coexisting with a novel topological phase transition that activates the vortex complexes. The evaluated boundaries of the topological phase transition in the parameter plane suggest that the topological phase under consideration is excited when a 'quantum' of the product between the anisotropy and the system size is available to the system.

Increasing the system size leads to a notable increase in the complexity of the collective organization. In contrast to the single vortex that characterizes small systems, larger systems exhibit a flexible morphogenetic flow, a complex spin-momentum locking pattern, that percolates throughout the entire system size. Although already observed in previously published initial studies [17, 18], the underlying mechanisms in these large-scale systems are still not fully understood and warrant further investigation.

The reasons of interest in the scenario put forward by this work are twofold. First is observing a novel topological phase in a classical dynamical system, unveiling a whole new class of non-equilibrium states. The robustness of the reported vortex complexes, along with their dissipationless transport and spin-momentum locking, are highly suggestive of a deeper theoretical underpinning.

Secondly, the emergence and self-organization of robust, non-local structures, which are sustained far from equilibrium by a dissipationless flow, and structured by topological constraints, provides a new perspective on biological organization. Biological systems are fundamentally defined by emergent, self-organized structures that operate far from thermodynamic equilibrium [22]. The emergence of flow and form from local interactions creating global (non-local) structures is a core principle of biological self-organization [23]. A principle reflected by the scenario displayed in this work, which moreover provides a mechanism for dissipationless flow that enables the system to persist efficiently out-of-equilibrium, sustained by a minimal energy input. This efficiency is critical, as – in physical systems – high dissipation releases heat (entropy production), which would otherwise limit the structures lifetime. Topological robustness provides resilience against noise and structural deterioration, granting the patterns a capacity for self-repair—another fundamental trait associated with living systems. Finally, the patterns show the emergence of a further, fundamental biological feature: *diversity*. This diversity is attributable, once again, to the topological character of those states, allowing for infinitely diverse, yet topologically equivalent, states composed of identical units, even under the same system conditions and parameters.

Far from being exhaustive, the present study only scratched the surface of a non-linear dynamical problem endowed with a surprising wealth of complex organizational dynamics.

**Supplementary Materials:** The following supporting information can be downloaded at: Video S1: title: Uniform type 1 pattern.
Video S2: title: Antivortex complex.
Video S3: title: Robustness against rupture
Video S4: title: Regime pattern for N=200, Δ=0,15.
Video S4: title: Regime pattern for N=500, Δ=0.1.
Video S4: title: Regime pattern for N=500, Δ=0.01.

**Author Contributions:** A.S. is the sole author of this manuscript and was responsible for all aspects of the reported research, including conception, theoretical development, analysis, and manuscript preparation.

**Funding:** This research received no external funding

**Institutional Review Board Statement:** Not applicable

**Informed Consent Statement:** Not applicable

**Data Availability Statement:** The author confirms that all data supporting the findings of this study are available within the article and its Supplementary Materials

**Acknowledgments:** The author thanks the University of Pavia for institutional support. The author is grateful for the personal dedication and tireless patience required to complete this work.

**Conflicts of Interest:** The author declares no conflicts of interest

## Abbreviations

The following abbreviations are used in this manuscript:

CW         Clockwise
CCW      Counterclockwise